\documentclass{pasj00}

\DeclareAbbreviation\AAHam{Astron. Abh. Hamburg. Sternw.}
\DeclareAbbreviation\AARv{Astron. Astrophys. Rev.}
\DeclareAbbreviation\an{Astron. Nachr.}
\DeclareAbbreviation\AcA{Acta Astron.}
\DeclareAbbreviation\Afz{Astrofizika}
\DeclareAbbreviation\AnTok{Tokyo Astron. Obs. Annals, Sec. Ser.}
\DeclareAbbreviation\Ap{Astrophysics}
\DeclareAbbreviation\ARep{Astron. Rep.}
\DeclareAbbreviation\ATel{Astronomer's Telegram}
\DeclareAbbreviation\ATsir{Astron. Tsirk.}
\DeclareAbbreviation\AcApS{Acta Astrophys. Sinica}
\DeclareAbbreviation\AstL{Astron. Letters}
\DeclareAbbreviation\BaltA{Baltic Astron.}
\DeclareAbbreviation\BASI{Bull. Astron. Soc. India}
\DeclareAbbreviation\BeSN{Be Star Newsletter}
\DeclareAbbreviation\GCN{GCN}
\DeclareAbbreviation\ibvs{Inf. Bull. Variable Stars}
\DeclareAbbreviation\JAD{J. Astron. Data}
\DeclareAbbreviation\JAVSO{J. American Assoc. Variable Star Obs.}
\DeclareAbbreviation\JBAA{J. British Astron. Assoc.}
\DeclareAbbreviation\LowOB{Lowell Obs. Bull.}
\DeclareAbbreviation\MitVS{Mitteil. Ver\"{a}nderl. Sterne}
\DeclareAbbreviation\MmSAI{Mem. Soc. Astron. Ita.}
\DeclareAbbreviation\Msngr{Messenger}
\DeclareAbbreviation\NewA{New Astron.}
\DeclareAbbreviation\NewAR{New Astron. Rev.}
\DeclareAbbreviation\OAP{Odessa Astron. Publ.}
\DeclareAbbreviation\Obs{Observatory}
\DeclareAbbreviation\PASA{Publ. Astron. Soc. Australia}
\DeclareAbbreviation\PAZh{Pis'ma AZh}
\DeclareAbbreviation\PhR{Phys. Rep.}
\DeclareAbbreviation\PVSS{Publ. Variable Stars Sect. R. Astron. Soc. New Zealand}
\DeclareAbbreviation\PZ{Perem. Zvezdy}
\DeclareAbbreviation\PZP{Perem. Zvezdy Pril.}
\DeclareAbbreviation\QJRAS{QJRAS}
\DeclareAbbreviation\RMxAA{Rev. Mexicana Astron. Astrof.}
\DeclareAbbreviation\RvMA{Reviews of Modern Astron.}
\DeclareAbbreviation\Sci{Science}
\DeclareAbbreviation\SvA{Soviet Astronomy}
\DeclareAbbreviation\SvAL{Soviet Astronomy Letters}
\DeclareAbbreviation\VeSon{Ver\"{o}ff. Sternw. Sonneberg}
\DeclareAbbreviation\VSOLJBul{VSOLJ Variable Star Bull.}
\DeclareAbbreviation\yCat{VizieR Online Data Catalog}
\DeclareAbbreviation\ZA{Z. Astrophys.}

\begin{document}
\SetRunningHead{M. Uemura, et al.}{Bayesian Separation of
  Polarization in Blazars}
\Received{2009/10/09}
\Accepted{2009/11/16}

\title{Bayesian Approach to Find a Long-Term Trend in
  Erratic Polarization Variations Observed in Blazars}

\author{
Makoto \textsc{Uemura}\altaffilmark{1},
Koji S. \textsc{Kawabata}\altaffilmark{1},
Mahito \textsc{Sasada}\altaffilmark{2},
Yuki \textsc{Ikejiri}\altaffilmark{2},\\
Kiyoshi \textsc{Sakimoto}\altaffilmark{2},
Ryosuke \textsc{Itoh}\altaffilmark{2},
Masayuki \textsc{Yamanaka}\altaffilmark{2}, 
Takashi \textsc{Ohsugi}\altaffilmark{1},\\
Shuji \textsc{Sato}\altaffilmark{3}, and
Masaru \textsc{Kino}\altaffilmark{3}}

\altaffiltext{1}{Hiroshima Astrophysical Science Center, Hiroshima University, Kagamiyama
1-3-1, \\Higashi-Hiroshima 739-8526}
\email{uemuram@hiroshima-u.ac.jp}
\altaffiltext{2}{Department of Physical Science, Hiroshima University,
Kagamiyama 1-3-1, \\Higashi-Hiroshima 739-8526}
\altaffiltext{3}{Department of Physics, Nagoya University, Furo-cho,
Chikusa-ku, Nagoya 464-8602}


%


\maketitle

\begin{abstract}

We developed a method to separate a long-term trend from observed
temporal variations of polarization in blazars using a Bayesian
approach.  The temporal variation of the polarization vector is
apparently erratic in most blazars, while several objects occasionally
exhibited systematic variations, for example, an increase of the
polarization degree associated with a flare of the total flux.  We
assume that the observed polarization vector is a superposition of
distinct two components, a long-term trend and a short-term variation
component responsible for short flares.  Our Bayesian model estimates
the long-term trend which satisfies the condition that the total flux
correlates with the polarized flux of the short-term component.  We
demonstrate that assumed long-term polarization components are
successfully separated by the Bayesian model for artificial data.  We
applied this method to photopolarimetric data of OJ~287,
S5~0716$+$714, and S2~0109$+$224. Simple and systematic long-term
trends were obtained in OJ~287 and S2~0109$+$224, while no such a
trend was identified in S5~0716$+$714. We propose that the apparently
erratic variations of polarization in OJ~287 and S2~0109$+$224 are due
to the presence of the long-term polarization component. The behavior
of polarization in S5~0716$+$714 during our observation period implies
the presence of a number of polarization components having a quite
short time-scale of variations. 

\end{abstract}

\section{Introduction}

Blazars are believed to be active galactic nuclei (AGN) with
relativistic jets pointing toward us (e.g. \cite{bla78blazar}).
Doppler-boosted non-thermal emission from the jet dominates from radio
to $\gamma$-rays in blazars.  The radio---optical emission is
dominated by synchrotron emission, and hence, highly
polarized. Polarimetric observations have been extensively performed
since they provide a clue for the structure of magnetic field in the
jet (e.g. \cite{ang80blazar_pol}; \cite{mea90blazar_pol}).  Another
important feature of blazars is a violent variability in a wide
range of time-scales from a few minutes to years
(e.g. \cite{huf92variation}).  Since the polarization degree (PD) and
angle (PA) are also variable with time, dense and long-term
polarimetric observations are essential to understand the variation
mechanism of blazars and the magnetic field structure in the jet. 

\citet{moo82bllac} conducted an intensive observation campaign for
BL~Lac with multi-longitude optical observatories over a week period,
and found that the polarization changes trace out a random walk in the
Stokes $QU$ plane.  In general, erratic variations have been observed
in polarization of most blazars (e.g. \cite{ang80blazar_pol}).  On the
other hand, 
systematic variations in PD and PA were also occasionally detected in
blazars, for example, flares associated with the increase of PD
(\cite{smi863c354}; \cite{tos98mrk421}; \cite{efi98on231}) and the
color variation correlating with the variation in PD
(\cite{cel07ao0235}).  \citet{mar08bllac} have recently reported a
smooth rotation of the polarization vector in BL~Lac, which
indicates a propagation of the emitting region through a helical
magnetic field in the jet.  Thus, both erratic and systematic
variations of polarization have been observed in blazars. 

\begin{figure}
 \begin{center}
 \FigureFile(80mm,80mm){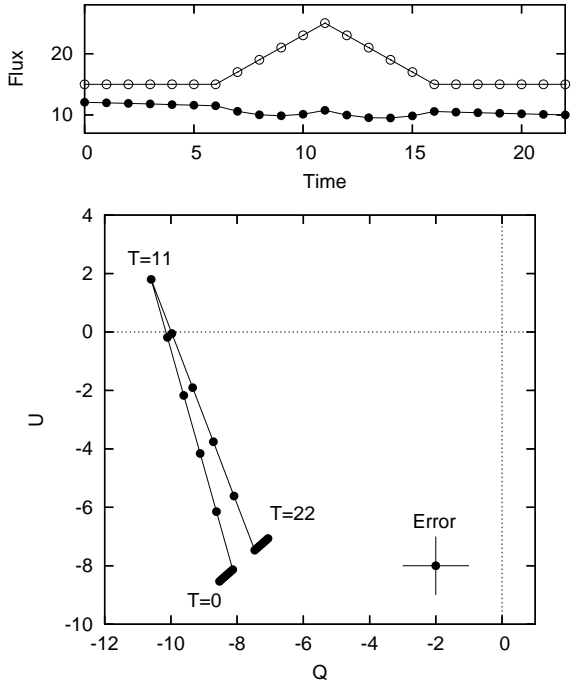}
 \end{center}
 \caption{Example of the short-term flare in the artificial data for
 our Bayesian model.  The upper panel shows the light curve of the
 total and polarized fluxes indicated by the open and filled circles,
 respectively.  The lower panel shows the temporal variation of
 polarization in the Stokes $QU$ plane.  A given error for the points
 is shown in the  right-lower corner of the panel.}\label{fig:example}
\end{figure}

The erratic variation of the polarization vector can be explained by a
scenario that a number of independent sources with randomly oriented
and strong polarizations blink (\cite{moo82bllac}; \cite{imp893c273};
\cite{jon85rotation}).  The systematic variations are, however,
difficult to be explained only by the random variation. Alternatively,
a number of systematic variations can be apparently overlooked in PD
and PA if short-term polarization variations are superimposed on a
long-term polarization trend.  A schematic example is shown in
figure~\ref{fig:example}.  The upper and lower panels show the light
curve of the total and polarized fluxes and the temporal variation of
the polarization vector in the Stokes $QU$ plane, respectively.  In
this example, the polarization vector is assumed to be a superposition
of two components; one shows a short-term variation associated with a
flare of the total flux, and the other is a long-term trend.  In the 
case that there is no long-term trend, PF correlates with the total
flux, independent of the PA of the short-term flare.  In the case that
the long-term trend has a significant PF as shown in the lower panel of
figure~\ref{fig:example}, however, the PF can show no positive
correlation with the light curve.  If the PA of the short-term flare
is random and observations are sparse, the variation of polarization
can be apparently erratic. 
 
The presence of the long-term component has been suggested
by several observations.  The short-term variability of blazars is
characterized by a combination of shots (\cite{huf92variation}).  A
typical time-scale of the shot was estimated to be $\sim 1$~d
(\cite{tak00mrk421}; \cite{kat01variation}).  In addition to these
short-term variations, it is well known that blazars show variations
with a time-scale of months--years.  It is possible that the long-
and short-term variation components have different polarization
features. \citet{mar02vlba} reported VLBI radio observations of
blazars, which show two or more components in polarization maps.  If
the polarization components in the radio range can be observed also in
the optical range, an observed optical polarization vector should be a
composition of those distinct radio polarization components since
the angular resolution of optical observations is much lower than that
of VLBI radio images.  \citet{hag02bllac} reported a long polarimetric
monitoring of BL~Lac in 1969--1991, and found the existence of the
preferred direction of the polarization vector.  They propose a model
that the polarization of BL~Lac has two kinds of sources, a stationary
component and a large number of randomly polarized components (also
see, \cite{bli09bllac}).  This stationary component could be
considered as a long-term trend if it varies gradually.

The two-component model shown in figure~\ref{fig:example} is one of
the simplest one among models with multiple polarization components.
It deserves further investigation if such a simple model can extract a
systematic trend from erratic variations of polarization in blazars. 
Here, we present an exploratory Bayesian approach to separate a
long-term trend from observed temporal variations of polarization in
blazars.  A description of the model and method is shown in the next
section.  We apply the model to observed photopolarimetric data of
blazars and discuss the implications of the results in section~3.  In
section~4, we discuss the validity of the model and results.  Finally,
we summarize our findings in section~5.

\section{Bayesian Model for the Separation of Long- and Short-Term
  Variation Components in Polarization}

\subsection{Model Description}

We assume following three conditions for the observed flux and
polarization in blazars.  i) An observed polarization vector is a
superposition of two distinct polarization vectors of long- and
short-term variation components.  ii) The short-term component is
responsible for observed variations of the total flux.  In other
words, the PF of the short-term component completely correlates with
the light curve of the total flux.  iii) Temporal variations of the
flux from the long-term component is small.  Condition ii) is
based on the fact that several flares of blazars are associated with
the increase of PF, as mentioned in section~1.  In our picture, all
flares are associated with PF flares although each variation is often
hidden in the observed PF because of the presence
of the stronger, long-term component.  Condition iii) is made just for
simplicity of calculation.  It is difficult to separate the short- and
long-term components simultaneously in both the light curve and the
polarization vector in our model.

Since Stokes parameters are additive, condition i) can be expressed
with Stokes parameters $Q$ and $U$ as follows: 
\begin{eqnarray}
\left\{
\begin{array}{l}
Q_{\rm obs} = Q_{\rm L} + Q_{\rm S},\\
U_{\rm obs} = U_{\rm L} + U_{\rm S},
\end{array}
\right.
\end{eqnarray}
where $(Q_{\rm obs},U_{\rm obs})$, $(Q_{\rm L},U_{\rm L})$, and 
$(Q_{\rm S},U_{\rm S})$ denote the linear polarization parameters of
the observed, long-, and short-term components, respectively.    The
short-term component $(Q_{\rm S},U_{\rm S})$ is determined when
$(Q_{\rm L},U_{\rm L})$ is given for $(Q_{\rm obs},U_{\rm obs})$.  PF, 
PD and PA are calculated as $\sqrt{Q^2+U^2}$, ${\rm PF}/I$, and
$\frac{1}{2}\arctan(U/Q)$, respectively.  

Bayesian statistics provides a method to estimate a posterior
probability density function (PDF) of model parameters from a
likelihood function defined by the model and the data and a prior PDF 
of the model parameters.   We develop a Bayesian model to estimate a
time-series of $Q$ and $U$ of the long-term trend, that is, 
$\mathbf{y}=\{Q_{{\rm L},i},U_{{\rm L},i}\}$ $(i=0-N)$.  
We use $N$ sets of photopolarimetric observations, 
$\mathbf{x}=\{Q_{{\rm obs},i},U_{{\rm obs},i}, I_{{\rm obs},i}\}$, 
where $I_{\rm obs}$ denotes the observed total flux. 
According to the Bayesian theorem, the posterior distribution of
$\mathbf{y}$ is calculated as:
\begin{eqnarray}
P(\mathbf{y}|\mathbf{x}) = \frac{L(\mathbf{y}|\mathbf{x})\pi(\mathbf{y})}{C}, 
\end{eqnarray}
where C is a constant for normalization of PDF.  Here,
$L(\mathbf{y}|\mathbf{x})$ and $\pi(\mathbf{y})$ are the likelihood
function and the prior distribution of $\mathbf{y}$, respectively.

The likelihood function, $L(\mathbf{y}|\mathbf{x})$, is defined 
based on our assumed condition ii) and iii).  The PF of the short-term 
component can be calculated as ${\rm PF}_{\rm S} = 
\sqrt{Q_{\rm S}^2+U_{\rm S}^2}$. 
We normalized $I_{\rm obs}$ and ${\rm PF}_{\rm S}$ using the,
so-called, ``Z-score'' transformation defined as
$a^\prime = (a-\bar{a})/\sigma_a$, where $a^\prime$, $a$, $\bar{a}$,
and $\sigma_a$ are the normalized and original parameter values, its
average, and standard deviation, respectively.  This normalization
procedure reduces the uncertainty of a possible contribution of the
long-term component to the total flux if the contribution is
time-independent as assumed in condition iii).  The likelihood
function is, then, defined with these normalized parameters, 
$I_{\rm obs}^\prime$ and ${\rm PF}_{\rm S}^\prime$, as follows: 
\begin{eqnarray}
L(\mathbf{y}|\mathbf{x})=
\prod_i \frac{1}{\sqrt{2\pi \sigma_{{\rm PF}^\prime,i}^2}} 
\exp\left\{
-\frac{(I_{{\rm obs},i}^\prime - {\rm PF}_{{\rm S},i}^\prime)^2}
{2\sigma_{{\rm PF}^\prime,i}^2} 
\right\},
\end{eqnarray}
where $\sigma_{{\rm PF}^\prime}$ is the error for 
${\rm PF}_{\rm S}^\prime$.  $\sigma_{{\rm PF}^\prime}$ is
approximately given by the error of the observed PF. 
We neglected the error of $I_{\rm obs}$ because photometric errors 
of the total fluxes are much smaller than the errors of polarization
parameters. $L(\mathbf{y}|\mathbf{x})$ reaches the maximum when the
profile of the light curve is the same as that of ${\rm PF}_{\rm S}$.  

The prior distribution, $\pi(\mathbf{y})$, plays a role in the control
of the time-scale of the long-term trend in our model.  Our model
requires that the variation time-scale of $(Q_{\rm L},U_{\rm L})$ is
longer than that of $(Q_{\rm S},U_{\rm S})$. A long-term 
variation definitely draws a smoother path than that of short-term
variations. A smooth curve can be described by a condition that a
sequence of the first difference of $\{Q_{{\rm L},i}\}$ and $\{U_{{\rm
    L},i}\}$ follows a standard normal distribution.  We define the
prior distribution as follows;  
\begin{eqnarray}
\pi(\mathbf{y})&=&\pi(\{Q_{{\rm L},i}\})\pi(\{U_{{\rm L},i}\}),\\
\pi(\{Q_{{\rm L},i}\})&=& \prod_i \frac{1}{\sqrt{2\pi w^2}}
\exp\left\{
-\frac{(Q_{{\rm L},i} - Q_{{\rm L},i-1})^2}{2w^2}
\right\},\\
\pi(\{U_{{\rm L},i}\})&=& \prod_i \frac{1}{\sqrt{2\pi w^2}}
\exp\left\{
-\frac{(U_{{\rm L},i} - U_{{\rm L},i-1})^2}{2w^2}
\right\}.
\end{eqnarray}
$\pi(\mathbf{y})$ has no physical meaning in our model.  It only
controls the smoothness of the long-term trend by a hyperparameter,
$w$.  The result obtained with this model evidently depends on $w$.
We determine an appropriate $w$ using so-called ``empirical Bayesian''
approach, by making a criterion for obtained results,
which are discussed in subsection~2.3 and subsection~4.1.  The
hyperparameter, $w$, has the same dimension as $Q$ and $U$.  Since $w$
corresponds to a standard deviation of a normal distribution as can be
seen in equations~(5) and (6), it is useful to use $w$ normalized by a
typical observation error of $QU$, $\sigma$;
\begin{eqnarray}
\sigma = \frac{\sum_{i=1}^N \sqrt{Q_{{\rm err},i}^2 + U_{{\rm err},i}^2}}{N}, 
\end{eqnarray}
where {$Q_{{\rm err},i}$} and {$U_{{\rm err},i}$} are the
observation error of $Q_{{\rm obs},i}$ and $U_{{\rm obs},i}$,
respectively.  In subsection~3.2, we discuss the dependence of the result
on $w/\sigma$.  

Thus, we can calculate $L(\mathbf{y}|\mathbf{x})$ and
$\pi(\mathbf{y})$ for $\mathbf{y}$ using the observations,
$\mathbf{x}$.  The posterior probability, $P(\mathbf{y}|\mathbf{x})$,
was then estimated based on equation~(2).  The estimation was
performed with the Markov Chain Monte Carlo (MCMC) method 
with the Metropolis algorithm (\cite{met53mcmc}).  The calculation
procedure of MCMC is as follows: At the $n$-th step of MCMC, we obtain 
$p_n = L_n\pi_n$ for $\mathbf{y}_n$.  We then randomly move to another
point in the parameter space of $\mathbf{y}$ by adding random values
drawn from a standard normal distribution which has a dispersion
chosen to efficiently sample the likelihood surface.  As a result, we
obtain $p_{n+1}$.  We count the $(n+1)$-th step if either
$p_{n+1}>p_n$ or $p_{n+1}/p_n$ is larger than a uniform random number
between 0 and 1.  Otherwise, we reject it and rework the $(n+1)$-th
step.  After discarding the first $10^4$ steps, we sample every 100
steps until the number of the sample becomes $10^5$.  This procedure
forms a set of $\mathbf{y}$, and we obtained ten sets of $\mathbf{y}$
with different initial values of $\mathbf{y}$.  We confirmed that no
significant difference was seen in each set of $\mathbf{y}$.  We
merged them, and finally, obtained the MCMC sample of $\mathbf{y}$.
The median and 68.3\% confidence intervals for $\mathbf{y}$ were
determined from this combined sample.

\subsection{Demonstration with Artificial Data}

\begin{figure*}
 \begin{center}
 \FigureFile(117mm,160mm){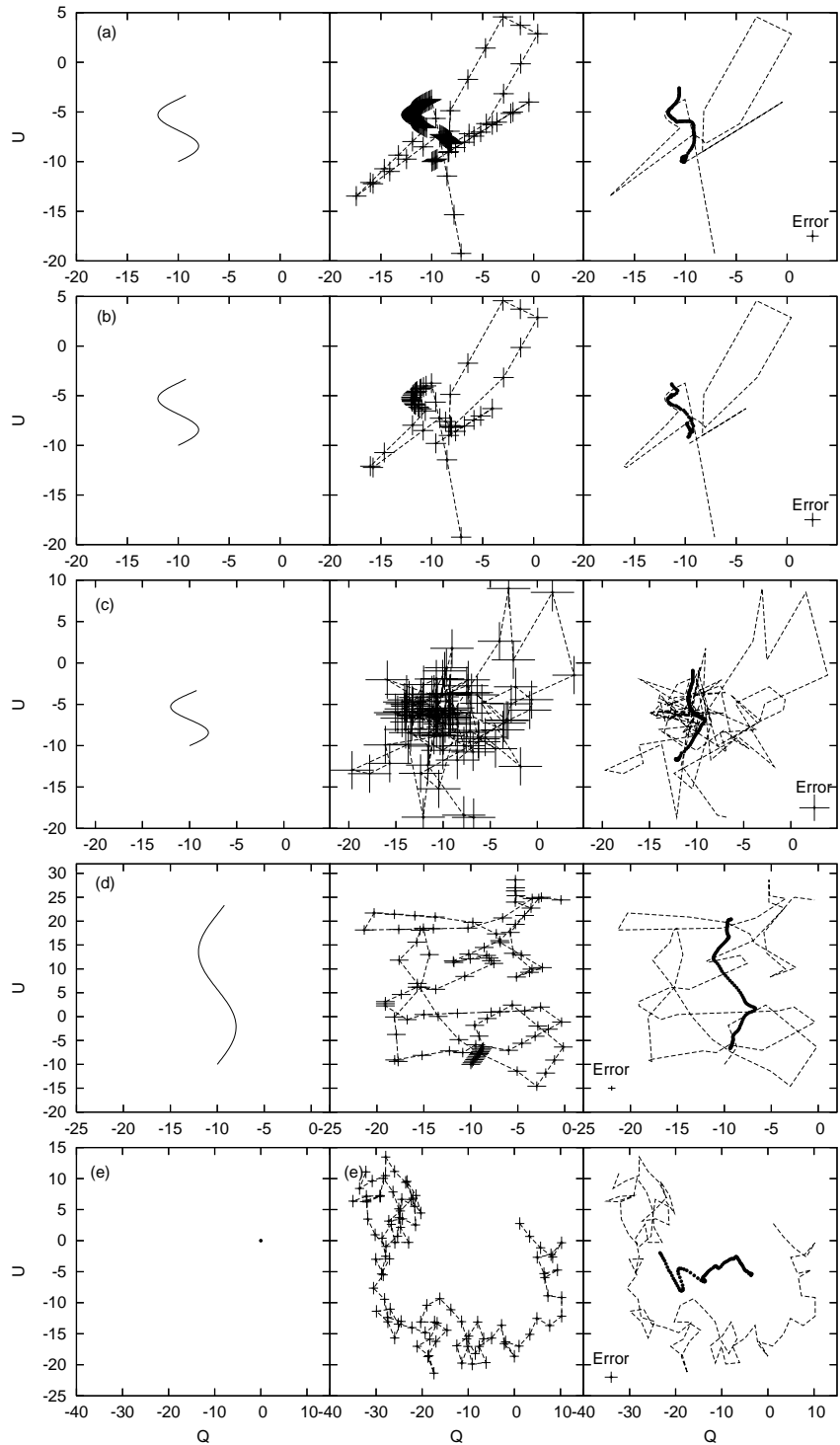}
 \end{center}
 \caption{Artificial polarization data and estimated long-term trends
 in the $QU$ plane.  Five sets of the data were used, as labeled (a),
 (b), (c), (d), and (e) (for the details of the data, see the text).
 The left, middle, and right panels show the assumed long-term trends,
 the short-term variations superimposed on the long-term trends, and
 the estimated long-term trends, respectively.  The paths of the $QU$
 variation of the data are also shown in the right panels with the
 dashed lines.  In Data~(e), no long-term trend was assumed.
 The Bayesian estimations of the long-term trend were performed with
 $w/\sigma=0.20$ in all cases.}\label{fig:simqu} 
\end{figure*}

\begin{figure*}
 \begin{center}
 \FigureFile(160mm,160mm){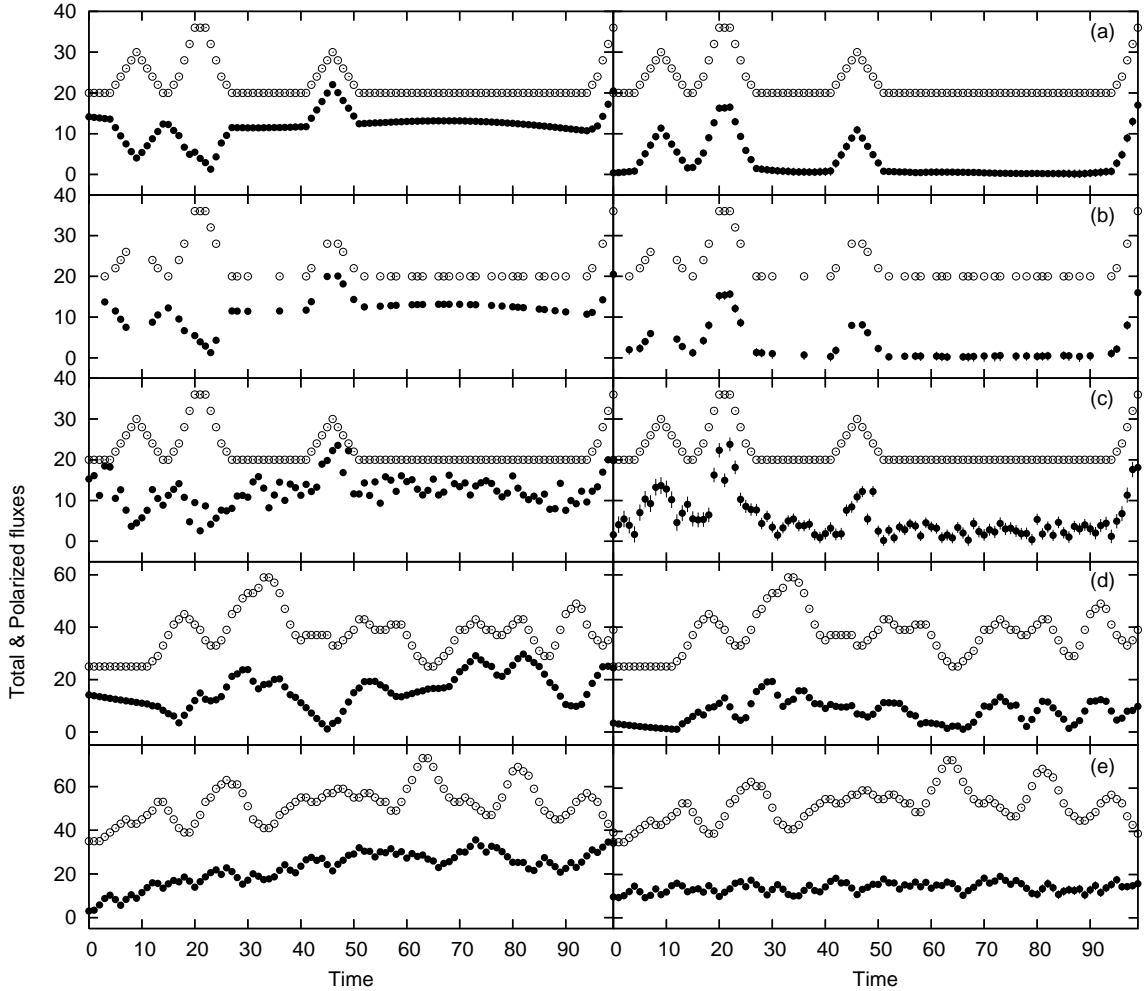}
 \end{center}
 \caption{Light curve of the total and polarized fluxes of the five
 sets of the artificial data labeled as (a), (b), (c), (d), and (e),
 indicated  by the open and filled circles, respectively.  In the
 right panels, the estimated long-term trends were subtracted from the
 polarized flux.}\label{fig:simlc}
\end{figure*}

In this section, we demonstrate how our Bayesian model works using
artificial data sets.  We generated five sets of artificial data of
temporal $QU$ variations and light curves.  Each data set consists of
100 sets of $Q$, $U$, and $I$ in a time-series from $t=0$ to $99$.
The duration and amplitude of the short-term flare were fixed in the
light curve and PF as $\Delta t = 10$, $\Delta I = 10$, and 
$\Delta {\rm PF}_{\rm S} = 10$.  Errors of $Q$ and $U$ were set to be
1.0.  We used simple linear functions for brightening and fading
phases of the flare. Figure~\ref{fig:example} shows an example of the
flare. Results of our analysis are independent of the flux out of the 
flare since the light curve is normalized in our procedure, as
mentioned above.  The short-term flares were randomly generated in the
time and PA.  The frequency of the flare was controlled with a
parameter, $\alpha$, which is a probability that a flare occurs at
each time. Overlaps of flares were allowed.  Figure~\ref{fig:simqu}
and \ref{fig:simlc} shows the variations in the $QU$ plane and the
light curves of the five sets of artificial data, labeled as (a), (b),
(c), (d), and (e), respectively. We assumed sine-curves for the
long-term trend, as depicted in the left panels of
figure~\ref{fig:simqu}.  The middle panels show $(Q_{\rm obs},U_{\rm
  obs})=(Q_{\rm L}+Q_{\rm S}, U_{\rm L}+U_{\rm S})$. 

We applied our Bayesian model to those artificial data.  The
hyperparameter, $w$, was fixed to $w/\sigma = 0.20$ in all cases.  
The estimated long-term trends were shown in the right panels of
figure~\ref{fig:simqu}, indicated by the filled circles.  Typical
errors of $(Q_{\rm L}, U_{\rm L})$ are also shown in each panel.  The 
temporal variations of ${\rm PF}_{\rm S}$ are shown in the right 
panels in figure~\ref{fig:simlc}.

Data (a) was generated with a low flare frequency of $\alpha=0.02$.
The second and forth flares apparently have larger amplitudes than the 
others in the light curve because two consecutive flares are
incidentally overlapped.  A sign of the assumed long-term trend can be 
seen in the middle panel of figure~\ref{fig:simqu} because of the low
flare frequency.  The observed PF apparently shows anti-correlations
with the light curve in the flares of $t<30$, while positive
correlations are seen in $t>30$.  This is because the
long-term trend has a significant polarization, and the polarization
vectors of the early flares, incidentally, directed to the origin of
the $QU$ plane.  As seen in figure~\ref{fig:simqu}, the Bayesian model
well reproduced the assumed long-term trend.  The corrected 
${\rm PF}_{\rm S}$ shows a clear correlation with the light curve in
all period, as can be seen in figure~\ref{fig:simlc}.  

Data (b) was generated from Data~(a) by random sampling.  We
``observed'' Data~(a) with an observation probability of 50~\%.  This
data is used in order to evaluate the Bayesian model for data taken 
with unequal time intervals, like typical ground-based observations of
blazars.  This experiment is important because the prior distribution
in our model assumes data observed
with equally-spaced intervals.  The estimated ${\rm PF}_{\rm S}$ still 
shows a clear correlation with the light curve.  The estimated
long-term trend is somewhat smoother than the assumed one, while it
successfully reproduces general features of the assumed trend. 

Data (c) is an example for data with a low signal-to-noise ratio
(S/N).  We generated $Q$ and $U$ of Data~(c) by adding a Gaussian
noise to those of Data~(a) with standard deviations of $Q$ and $U$ of
2.3.  This corresponds to S/N$=3.0$ for each polarization flare having
an amplitude of $\Delta {\rm PF}_{\rm S} = 10$.  The estimated
long-term component still shows a general trend of increasing $U$,
while the ``S''-shaped profile is weak.  The correlation between the
light curve and PF is improved even in the case that this long-term
trend is subtracted, as can be seen in figure~\ref{fig:simlc}.  This
is probably because the sub-structure of the assumed long-term trend
is negligible compared with the flares having a large amplitude in
this case.  We confirmed that the estimation of the long-term trend
was completely failed if the standard deviation of the Gaussian noise
was 3.5 (S/N$\sim 2$). In the case of a standard deviation of 1.8
(S/N$\sim 4$), the long-term trend was well reproduced.  Hence, our
model is validated if polarization flares are detected with
S/N$\gtrsim 3$ in the case of Data~(a).  In general, the allowable S/N
highly depends on the data itself.  It is possible that a higher S/N
is required for data with a higher flare-frequency. 

Data (d) shows more complicated temporal variations.  It was generated 
with a high flare frequency of $\alpha=0.30$, and a more prominent
long-term trend.  Almost no characteristic of the long-term trend is
seen in the middle panel of Data~(c) in figure~\ref{fig:simqu}, and
the path in the $QU$ plane is apparently erratic as generally observed
in blazars.  Nevertheless, the long-term trend was successfully
reproduced by the Bayesian model, as seen in the right panel of 
figure~\ref{fig:simqu}.  While there is no significant correlation
between the observed PF and the light curve, the correlation
coefficient between ${\rm PF}_{\rm S}$ and the light curve is
calculated to be $0.84_{-0.07}^{+0.05}$.  This demonstrates that the
Bayesian method can be a powerful tool for the analysis of apparently
erratic variations in polarization.  We note that residuals between
${\rm PF}_{\rm S}$ and the light curve are larger than those in the
case of Data~(a) and (b).  This indicates that the Bayesian model
possibly fails to reproduce the long-term trend in the case of a very
high flare frequency.

In Data (e), there is no association between the light curve and the
polarization parameters.  The light curve was generated in the same
way as in the other cases with $\alpha = 0.30$.  The variation of the
polarization vector was set to be a purely random walk in the $QU$
plane without a long-term trend.  The Bayesian model found a solution
of the long-term trend even in this case, as shown in the right panel
of Data~(d) in figure~\ref{fig:simqu}.  On the other hand, there is no
significant correlation between the estimated ${\rm PF}_{\rm S}$ and
the light curve.  This suggests that the model cannot extract any
meaningful long-term trend from Data~(d). 

\subsection{Evaluation of the Result Obtained from the Bayesian Model}
 
In the last section, we demonstrated that the Bayesian model
successfully reproduced the assumed long-term trends in the $QU$ plane
for the artificial data.  It can, however, generate a false long-term
trend even from the data in which no long-term trend was actually
present.  This is problematic for the application of this model to
real observations in which we have no information about the presence
of the long-term trend.  Hence, we need to make a criterion for the
validation of obtained results from the Bayesian model.  We also need
to determine the preferable range of the hyperparameter, $w$, because
the result depends on $w$.  In general, a hyperparameter in an
empirical Bayesian model can be estimated by maximizing marginal
likelihood. This standard method is, however, insufficient in our
model, as discussed in subsection~4.1.  Here, we make a specific
criterion for our model. 

First, the correlation of ${\rm PF}_{\rm S}$ and the light curve
should be significantly improved compared with that of the observed
PF and the light curve.  The significance of the difference in two
correlation coefficients is evaluated by $Z$-test.  Using two
correlation coefficients, $r_i$ ($i=1$ or 2), the test statistic, $Z$,
which has a standard normal distribution, is defined as follows:
\begin{eqnarray}
Z&=&(Z_1-Z_2)/\sigma,\\
Z_i &=& \frac{1}{2} \log \left( \frac{1+r_i}{1-r_i} \right),\\
\sigma&=&\sqrt{\sigma_{Z_1}^2+\sigma_{Z_2}^2},\\
\sigma_{Z_i} &=& \frac{1}{\sqrt{N_i-3}},
\end{eqnarray}
where $N_i$ is the number of samples.  We can conclude that the two
correlation coefficients are significantly different with a $>95$~\%
confidence level when $|Z|>1.96$.  In our Bayesian model, a smaller $w$
yields a smoother long-term trend, and hence results in a less
improvement in the correlation coefficients between ${\rm PF}_{\rm S}$
and the light curve.  As a result, a too small $w$ is discarded by
this procedure.  $|Z|$ were calculated to be 23.74, 17.41, 9.71, 6.86,
and 1.35 in the case of the artificial data (a), (b), (c), (d), and
(e) in the last subsection, respectively.  Thus, we conclude that the
long-term trend estimated for Data~(e) is a false result. 

Second, we need to check whether an estimated long-term trend actually
has a longer time-scale than short-term variations.  As mentioned
in subsection~2.1, a long-term component is expected to depict a
smooth path in the $QU$ plane, and adopted the prior distribution with
$w$.  In other words, the hyperparameter, $w$, is related to the
variation time-scale of the long-term trend in our model.  A path of
the long-term trend becomes more complicated and erratic in the case
of a larger $w$ (for example, see figure~\ref{fig:samplequ} in
subsection~3.2).  With an extremely large $w$, the variation amplitudes and
time-scales of $Q_{\rm L}$ and $U_{\rm L}$ can be comparable to those
of $Q_{\rm S}$ and $U_{\rm S}$. Thus, $w$ should be restricted to be
small enough to assure that the estimated $Q_{\rm L}$ and $U_{\rm L}$
actually exhibit ``long''-term variations. We evaluate it using a
travel distance in the $QU$ plane of both long- and short-term
components.  The travel distance, $d$, is defined for a set of
$\{Q_i,U_i\}$ as follows:  
\begin{eqnarray}
d = \sum_{i=1}^{N-1} \sqrt{(Q_{i+1}-Q_i)^2+(U_{i+1}-U_i)^2} .
\end{eqnarray}
In this paper, the estimated $Q_{\rm L}$ and $U_{\rm L}$ are
acceptable in the case that the ratio of $d$ for long- and short-term
components, $d_{\rm L}$ and $d_{\rm S}$, satisfies the following
condition: 
\begin{eqnarray}
R = \frac{d_{\rm L}}{d_{\rm S}} < 0.10.
\end{eqnarray}
We call $R$ as the ``distance ratio''.  In conjunction with the 
first criterion about the correlation coefficient, we can restrict $w$
in a narrow range.  

Results obtained by the Bayesian model are quantitatively evaluated
by the above criterion.  It is, however, possible that a false
long-term trend would be extracted from observations even the result
satisfies the above criterion.  Hence, results from the Bayesian
model should be evaluated carefully.  In addition to the above
criterion, we can check the validity of results with their qualitative
features. We expect that a  long-term trend exhibits not a random
walk, but a systematic motion in the $QU$ plane.  A long-term trend
depicting a very complicated path in the $QU$ plane may merely indicate
that there is no systematic long-term trend in the observed $QU$.  We
should also suspect a result highly sensitive to a small change in
$w$.

\section{Results: Application for Observed Photopolarimetric Data of
  Blazars} 

\subsection{Data Description}

\begin{figure*}
 \begin{center}
 \FigureFile(170mm,170mm){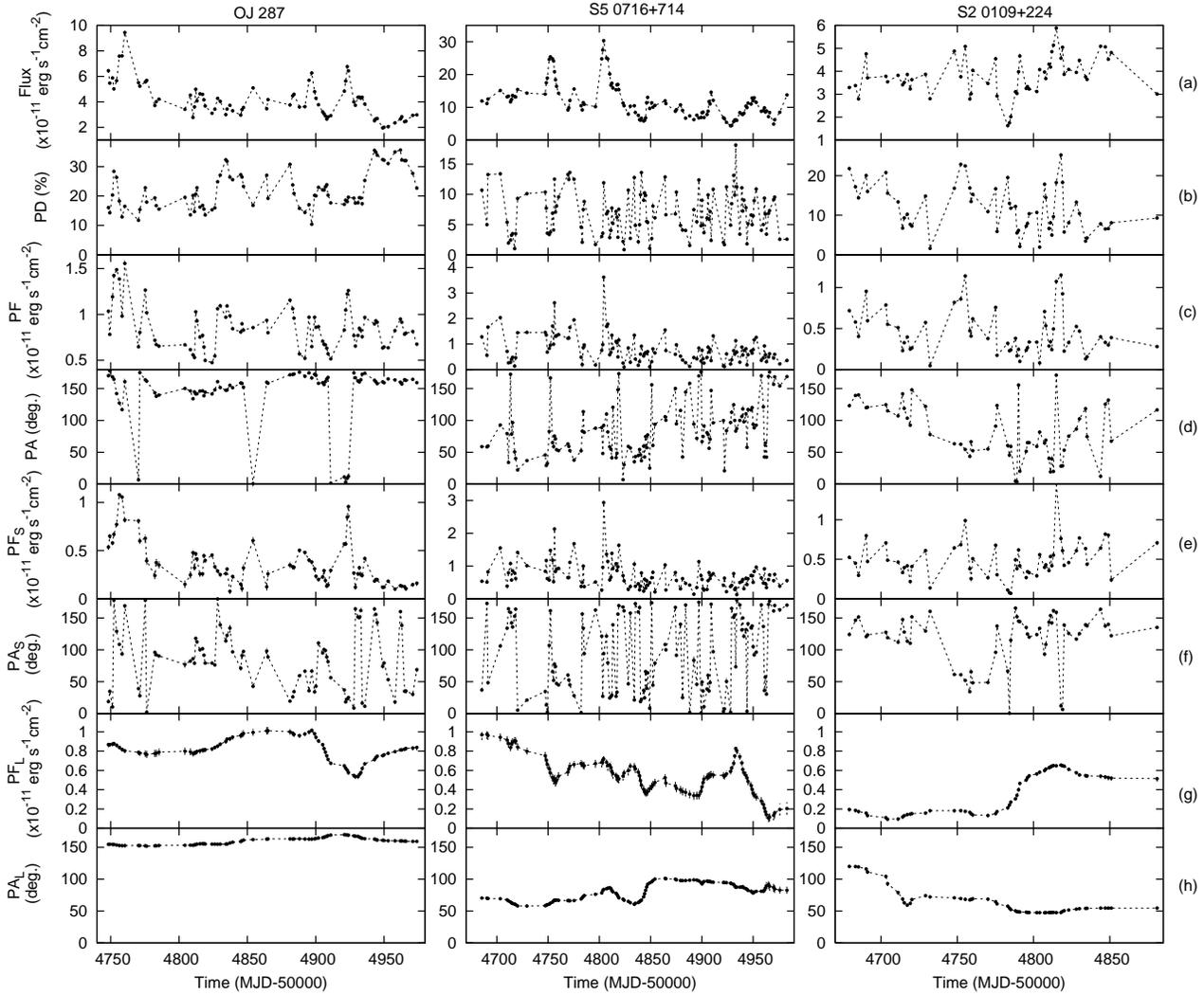}
 \end{center}
 \caption{Temporal variations of the observed and estimated
 parameters.  The left, middle, and right panels show the results of
 OJ~287, S5~0716$+$714, and  S2~0109$+$224, respectively, as indicated 
 at the top of the panels.  From top to bottom, the panels
 show (a) the observed light curve, (b) the observed polarization
 degree, (c) the observed polarized flux, (d) the observed
 polarization angle, (e) the estimated polarization flux of the
 short-term component, (f) the estimated polarization angle of the
 short-term component, (g) the estimated polarization flux of the
 long-term component, and (h) the estimated polarization angle of the
 long-term component.  The observation errors and 1-$\sigma$
 confidence level of the estimated parameters are also indicated,
 while most of them are comparable with the symbol
 size.}\label{fig:samplelc} 
\end{figure*}

We applied the Bayesian model to observed polarimetric data of
blazars.  The data were obtained with TRISPEC attached to the
``Kanata'' 1.5-m telescope at Higashi-Hiroshima Observatory during our 
photopolarimetric observation campaign for blazars in 2008--2009
(\cite{TRISPEC}; \cite{uem08kanata}).  A full description about the
observation and the data reduction will be published in a forthcoming
paper.  In this paper, we focus on how our model works in the real
data. 

\begin{figure*}
 \begin{center}
 \FigureFile(160mm,160mm){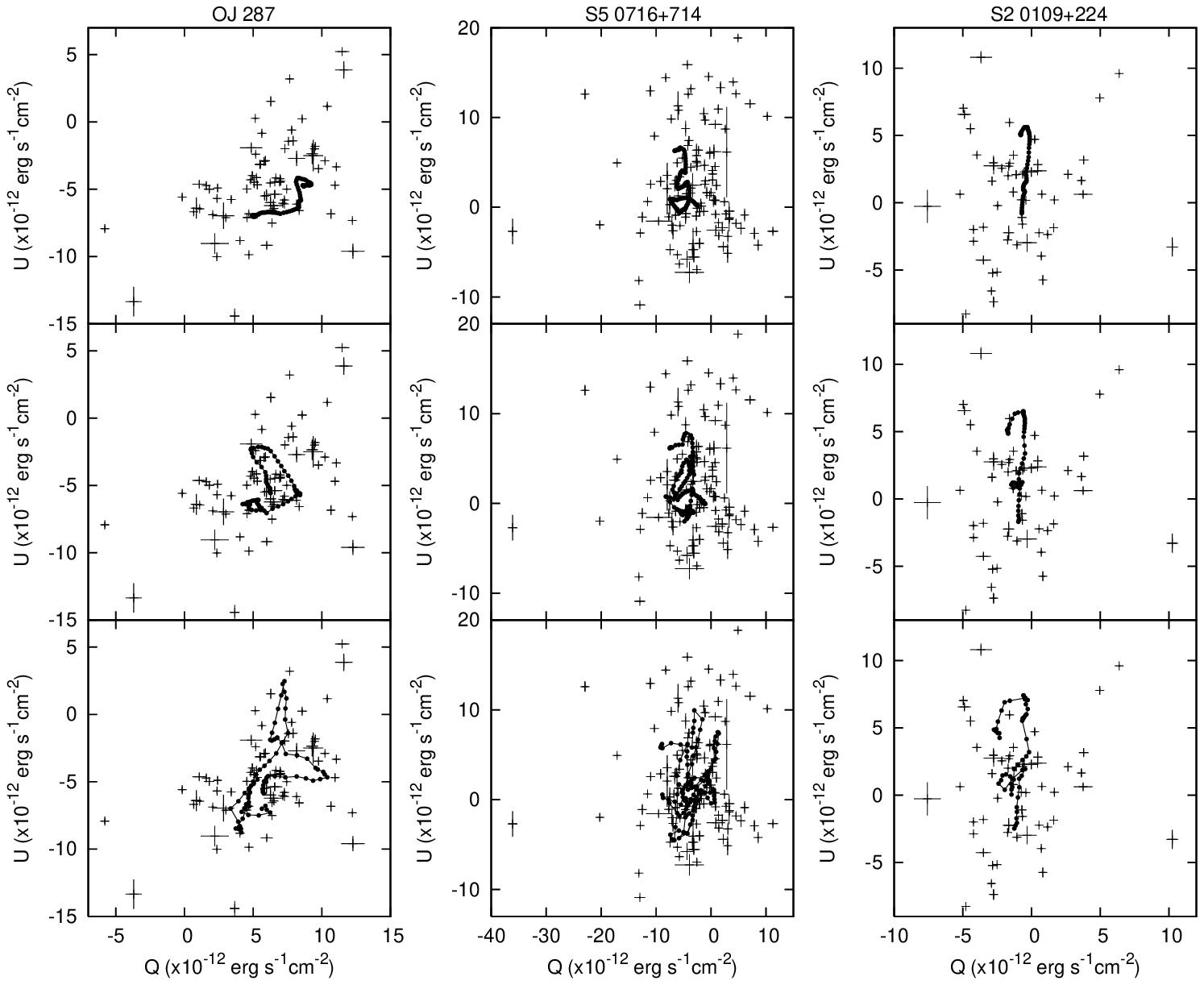}
 \end{center}
 \caption{Observed $QU$ variations and estimated long-term trends.
 The left, middle, and right panels show the data of OJ~287,
 S5~0716$+$714, and S2~0109$+$224, respectively, as indicated at the
 top of the panels.  The long-term trends were estimated with
 $w/\sigma=0.10$, 0.20, and 0.50 in the top, middle, and bottom panels, 
 respectively.  The observations and errors were indicated by the
 crosses.  The long-term trends were indicated by the filled circles.
 The errors for each points of the estimated long-term trends are
 typically less than 0.3, 1.5, and $0.3\times 10^{-12}\,{\rm
 erg}\,{\rm s}^{-1}\,{\rm cm}^{-2}$ for OJ~287, S5~0716$+$714, and
 S2~0109$+$224, respectively.}
\label{fig:samplequ}
\end{figure*}

We used the $V$-band photopolarimetric data of OJ~287, S5~0716$+$714,
and S2~0109$+$224.  The observations of these three objects were
performed in 79 nights between Oct 2008 and May 2009, 118 nights
between Aug 2008 and May 2009, and 56 nights between Jul 2008 and Feb
2009, respectively.  The light curves, PD, PF, and PA are shown in 
panels~(a), (b), (c), and (d) in figure~\ref{fig:samplelc},
respectively.  Variations in the $QU$ plane are shown in
figure~\ref{fig:samplequ}. 

OJ~287 exhibited short-term flares having a time-scale of days
superimposed on a gradual fading trend in the light curve.  In
contrast, PD remained high with 10--20\% during the observation
period, occasionally showing flares reaching maxima of $\sim 30$\%.  
The object almost stayed in the fourth quadrant in the $QU$ plane.
This behavior implies the presence of two polarization components,
that is, one for a long-term trend pointing to the fourth quadrant,
and the other for short-term flares.  

Short-term variations in S5~0716$+$714 had features similar to that
in OJ~287, while the behavior of PD is totally different.  The PD of
S5~0716$+$714 rapidly fluctuated between 0 and 15\%.  No systematic
variation correlating with the total flux can be seen in the
PD variation, except for a possible anti-correlation during a flare
around MJD~54750.  The temporal variation in the $QU$ plane is very
erratic. 

S2~0109$+$224 experienced a flare clearly associated with an
increase of PD around MJD~54810--54820.  Except for this flare,
no clear correlation can be seen between the light curve and 
PD.  The PA apparently shows a systematic long-term trend; 
a gradual decreasing trend until MJD~54800, followed by an increasing  
trend, while short-term variations occasionally disturb the trend.
This behavior in PA indicates the presence of long- and short-term
polarization components, as in OJ~287.  

\subsection{Bayesian Analysis of the Polarization Data}

The Bayesian estimation of the long-term trend was performed for these
three objects with $w/\sigma=0.10$, 0.20, 0.30, 0.40, and 0.50.  The
different $w/\sigma$ were used in order to check the dependency of
results on $w$. Figure~\ref{fig:samplequ} shows the estimated
long-term trends in the $QU$ plane indicated by the filled circles.
The top, middle, and bottom panels show the results with
$w/\sigma=0.10$, 0.20, and 0.50, respectively.  The long-term trends
draw more complicated paths in the $QU$ plane with larger $w/\sigma$,
as expected.  The test statistics, $|Z|$, and distance ratio, $R$,
defined in subsection~2.3 for those results are summarized in
table~\ref{tab:dist}.  Only the cases of $w/\sigma=0.20$ for OJ~287
and S2~0109$+$224 are acceptable in our criterion; $|Z|>1.96$ and
$R<0.10$.  Temporal variations of PF and PA of the short-term
components are shown in panels~(e) and (f) of
figure~\ref{fig:samplelc}, respectively.  Those of the long-term
trends are in panels~(g) and (h).  Those panels show the results with
$w/\sigma=0.20$. 

\begin{table}
 \caption{Test statistics, $|Z|$, and Distance ratio, $R$, in the case 
 of OJ~287, S5~0716$+$714, and S2~0109$+$224.}\label{tab:dist}
 \begin{center}
 \begin{tabular}{cccc}
 \hline
            & OJ~287 & S5~0716$+$714 & S2~0109$+$224\\
 $w/\sigma$ & \multicolumn{3}{c}{$|Z|$, $R$}\\
 \hline
 0.10 & 1.08, 0.04 & 0.33, 0.03 & 1.66, 0.04 \\ 
 0.20 & 2.85, 0.09 & 0.40, 0.06 & 2.27, 0.07 \\
 0.30 & 3.97, 0.14 & 1.39, 0.09 & 2.26, 0.10 \\
 0.40 & 4.66, 0.18 & 2.25, 0.13 & 2.75, 0.11 \\
 0.50 & 4.78, 0.21 & 2.38, 0.15 & 2.83, 0.16 \\
 \hline
 \end{tabular}
 \end{center}
\end{table}

The long-term trend in S5~0716$+$714 draws a rather erratic path in
the $QU$ plane even with $w/\sigma=0.10$.  The two-component model is 
probably inadequate to explain the observed $QU$ variation in
S5~0716$+$714. The $QU$ variation in S5~0716$+$714 may be caused 
by a number of polarization components having a variation time-scale
of $< 1$~d (\cite{moo82bllac}; \cite{jon85rotation}). In this case,
the time resolution of our observation could be too low to detect
systematic variations of polarization.  Rapid variations having a
time-scale of less than hours have been actually reported in
S5~0716$+$714 (\cite{sta06s50716}; \cite{sas08s50716}).

As can be seen from figure~\ref{fig:samplelc} and \ref{fig:samplequ},
the PA of the long-term trend in OJ~287 gradually increases until
MJD~54920, and then decreases with time.  In other words, the
polarization vector apparently oscillates between a narrow range
of PA of $150^\circ$--$170^\circ$ smoothly in the observation period.
This feature of the long-term 
trend can be seen both in the case of $w/\sigma=0.10$ and $0.20$,
which indicates that it is a stable feature.  The PF of the long-term
trend shows variations with a small amplitude of a factor $\sim 2$.
The long-term trend estimated in OJ~287 would, thus, be an ideal
example for our two-component model.

In the case of S2~0109$+$224, the PF of the long-term trend first
decreased until $\sim$MJD~54710, and then, the object stagnated near 
the origin of the $QU$ plane for 40~d.  After that, the PF increased
with the opposite direction to the early trend in the $QU$ plane, as
shown in figure~\ref{fig:samplequ}.  This component turned to decrease
again at last.  This feature can be commonly seen in all
cases of $w/\sigma$, while sub-structures become prominent in the case of
$w/\sigma=0.50$.  This behavior of the long-term trend may be
interpreted as two distinct components; one which kept decaying
until $\sim$MJD~54710, and the other which became prominent later.  
We note that the PF of the long-term trend shows a large
variation with a factor of $\sim 7$.  Since our two-component model
assumes no flux variation for the long-term trend, the large variation
amplitude of PF might violate this assumption.  The problem can be
reconciled only when the long-term trend exhibits a major variation in 
PD with a constant total flux.

\begin{figure}
 \begin{center}
 \FigureFile(75mm,75mm){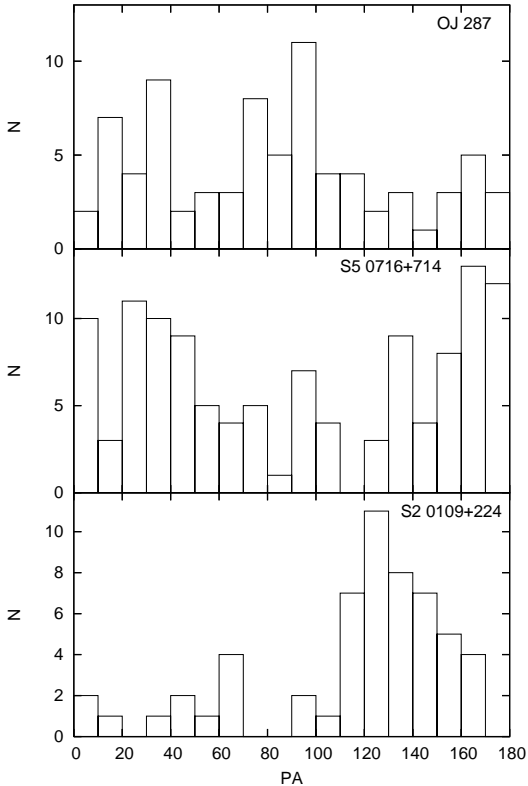}
 \end{center}
 \caption{Histogram of PA of the short-term component obtained with
 $w/\sigma=0.20$.  The top, middle, and bottom panels show the
 histograms for OJ~287, S5~0716+714, and S2~0109$+$224,
 respectively.}\label{fig:pahist}  
\end{figure}

The most notable feature in the estimated short-term component is
nonuniform distributions in PA in S2~0109$+$224 and possibly also 
OJ~287. Figure~\ref{fig:pahist} shows the histograms of the PA of the 
short-term components.  The figure also includes the distribution in
S5~0716$+$714 just as a reference.  We tested the
nonuniformity of the distributions using the Kolmogorov-Smirnov (KS)
test.  As a result, KS probabilities were calculated to be 0.09, 0.05,
and $<0.01$ in the case of OJ~287, S5~0716$+$714, and S2~0109$+$224,
respectively.  Therefore, we confirmed that the nonuniformity is
statistically significant in S2~0109$+$224 with a $>95$~\% confidence
level (,~or KS probability $< 0.05$).  This indicates that the PA of
the short-term component is not completely random, and hence implies
that the source of the short-term variation preferentially localizes
in an area where the direction of the magnetic field is possibly fixed.

Finally, we note that small $w$ were actually rejected in all cases of 
the three objects because no significant improvement was achieved in
the correlation between the total and polarized flux, as shown in
table~1.  This means that too simple long-term trends are insufficient
for our model.  The most simple long-term component would be
considered as averages of $Q$ and $U$ during the observation period.
If the correlation would be improved with the differential
polarization flux from the average $(Q,U)$, the Bayesian model was not
required.  However, we confirmed that the correlation is not
significantly improved with those differential polarization flux from
the averages; $|Z|=1.22$, 0.40, and 1.27 for OJ~287, S5~0716$+$714, and
S2~0109$+$224, respectively.  Our Bayesian model, therefore, has an
advantage to extract long-term trends from polarization variations,
compared with the simple correction with the average $(Q,U)$. 

\section{Discussion}

\subsection{Estimation of the Hyperparameter, $w$, by Maximizing
 Marginal Likelihood}

\begin{table}
 \caption{Marginal likelihood, $M(w)$, calculated with different
 $w/\sigma$.}\label{tab:mar} 
 \begin{center}
 \begin{tabular}{cccc}
 \hline
 $w/\sigma$ & OJ~287 & S5~0716$+$714 & S2~0109$+$224\\
     & \multicolumn{3}{c}{$\times 10^3$} \\
 \hline
 0.10 & 2.06 & 2.86 & 0.97\\
 0.50 & 2.96 & 3.96 & 1.92\\
 1.00 & 3.20 & 4.44 & 2.16\\
 2.00 & 3.21 & 4.61 & 2.21\\
 3.00 & 2.99 & 4.53 & 2.22\\
 5.00 & 2.60 & 4.15 & 2.17\\
 \hline
 \end{tabular}
 \end{center}
\end{table}

As mentioned in subsection~2.3, we made the criterion to determine 
an appropriate value of the hyperparameter, $w$, in our Bayesian
model. In general, a hyperparameter of an empirical Bayesian model can
be estimated by maximizing a marginal likelihood.  The marginal
likelihood, $M$, is equivalent with the constant, $C$, in
equation~(2), defined as:
\begin{eqnarray}
M(w) = \frac{L(\mathbf{y}|\mathbf{x})\pi(\mathbf{x}|w)}
            {P(\mathbf{y}|\mathbf{x})}.
\end{eqnarray} 
We calculated $M(w)$ based on the method described in \citet{chi01ml}.
The model parameter, $\mathbf{y}$, can be arbitrary since $M(w)$ is
independent of $\mathbf{y}$.  The numerator of equation~(14) was
calculated with the estimated best parameters of $\mathbf{y}$.
The denominator can be estimated by the Monte Carlo integration 
drawing a sample from $P(\mathbf{y}|\mathbf{x})$ which were obtained
by our Bayesian analysis.  The results are summarized in
table~\ref{tab:mar}.

$M(w)$ takes the maximum with $w/\sigma =2.00$--$3.00$ for all cases
of the three objects.  A long-term component estimated with such a 
large $w/\sigma$ definitely draws a quite complicated path in the
$QU$ plane, as can be seen from figure~\ref{fig:samplequ}.  As a
result, it is not acceptable in our criterion about the distance
ratio, $R$.  

The ratio of two $M(w)$ is called the ``Bayes factor'' (BF), which used 
to evaluate the models.  The BF are calculated to be $\sim 1.6$,
$\sim 1.6$, and $\sim 2.3$ with $M(0.10)$ and the maximum of $M$ for
OJ~287, S5~0716$+$714, and S2~0109$+$224, respectively.  These BF are
too small to conclude that the model with $w/\sigma =2.00$--$3.00$ is
significantly better than that with $w/\sigma =0.10$ (\cite{kas95bf}).  

Thus, the maximization of $M(w)$ is not suitable to determine $w$ in
our model.  This is mainly because the prior distribution is not the real
distribution for $Q_{\rm L}$ and $U_{\rm L}$.  Another form of the
prior distribution may be more suitable for the case of polarization
variations in blazars.  The practical criterion defined in subsection~2.3
is enough for our exploratory model in this paper.  

\subsection{Application in the ($Q/I,U/I$) Plane}

\begin{figure*}
\begin{center}
\FigureFile(160mm,160mm){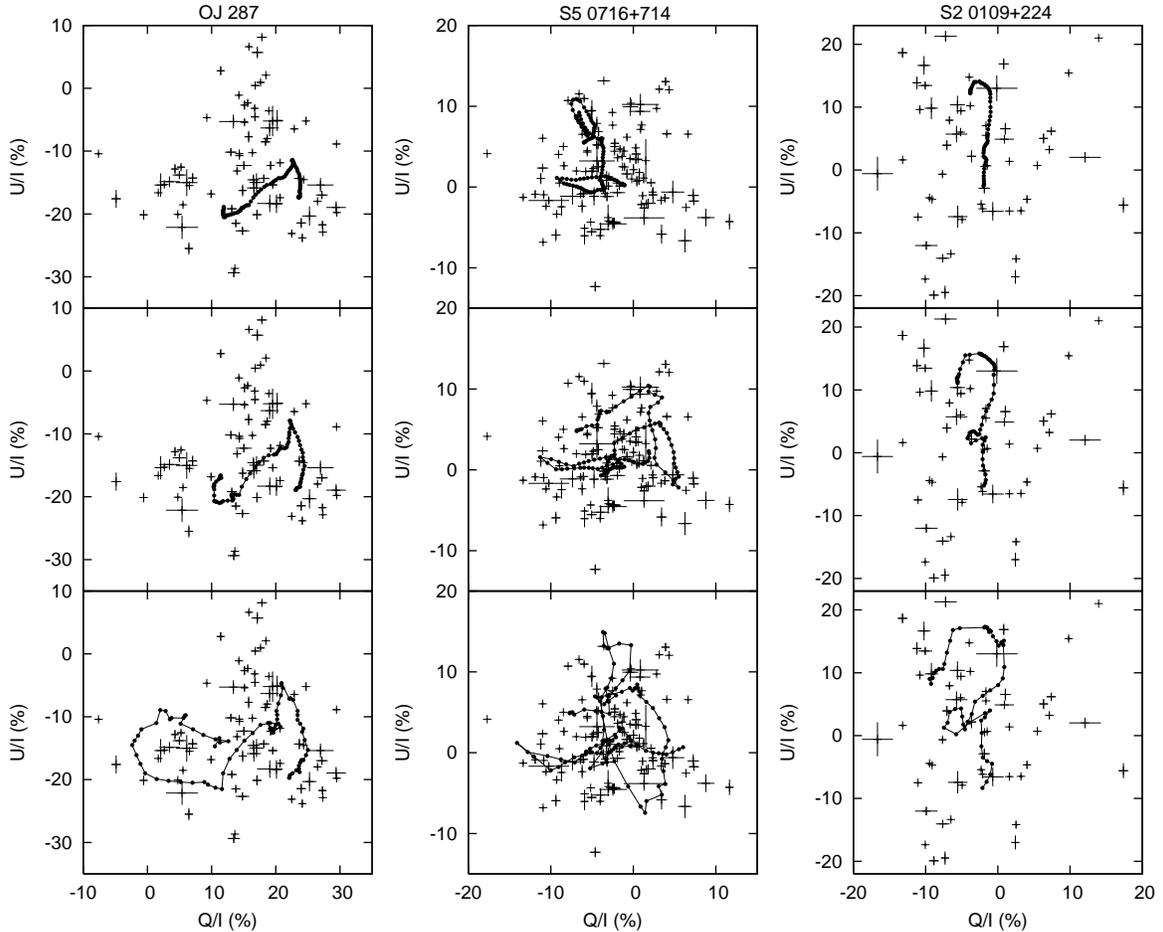}
\end{center}
\caption{Same as figure~\ref{fig:samplequ}, but for $Q/I$ and
 $U/I$.  The errors for each points of the estimated long-term
 trends are typically less than 1.0, 1.3, and 0.9~\% for OJ~287,
 S5~0716$+$714, and S2~0109$+$224, respectively.}\label{fig:samplequd}  
\end{figure*}

Our Bayesian model can find a long-term trend in the $QU$ plane.  It
can be also applied in the $(q,u)=(Q/I,U/I)$ plane, while strong
limitations are imposed for the application.  In our two-component
model, the observed $(q,u)$ is described as: 
\begin{eqnarray}
q = \frac{Q_{\rm L} + Q_{\rm S}}{I_{\rm total}}, 
u = \frac{U_{\rm L} + U_{\rm S}}{I_{\rm total}}\\
I_{\rm total} = I_{\rm L} + I_{\rm S}.
\end{eqnarray}
The two polarization components, namely, 
$(q_{\rm L},u_{\rm L})=(Q_{\rm L}/I_{\rm L},U_{\rm L}/I_{\rm L})$ and 
$(q_{\rm S},u_{\rm S})=(Q_{\rm S}/I_{\rm S},U_{\rm S}/I_{\rm S})$, 
can be separated in the $qu$ plane when $q_{\rm L}$ and $u_{\rm L}$
are much smaller than $q_{\rm S}$ and $u_{\rm S}$
(e.g. \cite{mes97polari}).  If the PD of the long-term trend is not
negligible, the polarization components cannot be separated without
the estimation of $I_{\rm L}$ and $I_{\rm S}$.  

In the case that the PD of the long-term trend is large, the
application of the Bayesian model in the $qu$ plane means the estimate
of $(Q_{\rm L}/I_{\rm total}, U_{\rm L}/I_{\rm total})$ that lead to a
positive correlation between the light curve and the modified PD
calculated from $(Q_{\rm S}/I_{\rm total}, U_{\rm S}/I_{\rm total})$.
Since $I_{\rm L}$ is assumed to be time-independent in our model, the
temporal behavior of the estimated $(Q_{\rm S}/I_{\rm total}, U_{\rm
  S}/I_{\rm total})$ is probably analogous to that of $(Q_{\rm
  S}/I_{\rm S}, U_{\rm S}/I_{\rm S})$.  In this sense, the Bayesian
analysis in the $qu$ plane is not completely nonsense, and can still
provide meaningful results.  A result in the $qu$ plane would be
noteworthy, particularly if the long-term trend in the $qu$ plane
draws an analogous path to that in the $QU$ plane.  

Figure~\ref{fig:samplequd} is the same as figure~\ref{fig:samplequ},
but for the $qu$ plane. We found that the long-term trends estimated
with $w/\sigma =0.20$ are quite analogous to those in
figure~\ref{fig:samplequ} in the case of S2~0109$+$224.  The long-term
trend in OJ~287 also exhibits a common feature with that in the $QU$
plane in terms of the oscillation of the polarization vector within a
narrow range of PA.  The similarity between the long-term trends in
the $QU$ and $qu$ planes implies that the short-term flares were
associated with the increases of PD in OJ~287 and S2~0109$+$224. 
The short-term flares may originate from a region where the direction
of the magnetic field is more aligned than the whole emitting region.

\subsection{Future Studies}

We showed that the polarization vector in OJ~287 and S2~0109$+$224 can
be separated into two components, the long-term trend and the
short-term variation component.  The result, however, provides no
clear evidence for the existence of the two components in these
objects.  Our Bayesian model just gives us one of possible views for
temporal variations of polarization in blazars. This view deserves to
be discussed since simple and systematic variations can be extracted
from apparently erratic variations. On the other hand, further
investigations are needed to confirm the two-component view.  

One of the most direct ways to obtain evidence for the two-component
scenario would be VLBI radio observations.  \citet{mar02vlba} reported the
presence of multiple polarization components even within radio cores in
blazars.  VSOP-2 is a space VLBI project, which will resolve sub-pc 
regions around the central blackhole of AGN (\cite{VSOP2}).  VSOP-2
will reveal the detailed structure of the polarization components in
the radio core.  A part of the radio polarization
components might exhibit temporal variations synchronized with the
optical polarization components resolved by our Bayesian
model.  If this is the case, our Bayesian model, for the first time,
provides a tool to investigate the temporal evolution of each
polarization component observed both in the optical and radio regimes. 

Our model assumes that the PF of the short-term component correlates
with the light curve of the total flux (condition ``ii)'' in
section~2.1).  In other words, we assumes the total flux exhibits a
flare whenever a short-term polarization flare occurs.  This
assumption gives the condition that the application of our Bayesian
model is validated.  Our model definitely yields a false result if it
applies to, for example, short-term polarization flares without
significant variations of the total flux.  If such variations are
typical in a blazar, our model is not validated for it.  Our
assumption should, hence, be evaluated case by case for objects.  In
our picture, a short-term flare with a larger amplitude is expected to
show a better correlation with polarization because the specific 
polarization-feature of a smaller flare would be readily diluted by
another small flares.  Hence, the polarization behavior of short and
large flares would provide a key for the validation of our assumption.

\citet{mar08bllac} reported a rotation of the optical polarization
vector in BL~Lac and interpreted it as a shift of the emitting region
through a helical magnetic field in the jet.  Such a rotation event
can have been missed if the observed polarization has a long-term
trend as considered in our model.  Our Bayesian model would provide a
tool to search the rotation events of the polarization vector both in
the observed, long- and short-term components.  No systematic rotation
event can be seen in the long- and short-term components in the three
objects, as shown in subsection~3.2. 

\section{Summary}

We developed a Bayesian model to extract a long-term trend from
apparently erratic variations in polarization of blazars.  Our
Bayesian model successfully resolved the long- and short-term
components in the artificial data.  We applied this model
to the photopolarimetric data of OJ~287, S5~0716$+$716, and
S2~0109$+$224.  Simple and systematic long-term trends were obtained
in OJ~287 and S2~0109$+$224, while no meaningful trend was identified
in S5~0716$+$716.  We propose that all short-term flares were
associated with the increase of the polarized flux in OJ~287 and
S2~0109$+$224.  The polarization variation in S5~0716$+$716 may be
explained by a random variation caused by a number of polarization
components having a quite short time-scale of variations. 

We acknowledge useful discussions with Taichi Kato and Takuya
Yamashita.  This work was partly supported by a Grand-in-Aid from the
Ministry of Education, Culture, Sports, Science, and Technology of
Japan (19740104).


\end{document}